# Multiscale principal component analysis

A A Akinduko and A N Gorban

Mathematics Department, University of Leicester, Leicestershire, LE1 7RH, UK

E-mail: aaa78@le.ac.uk and ag153@le.ac.uk

**Abstract**. Principal component analysis (PCA) is an important tool in exploring data. The conventional approach to PCA leads to a solution which favours the structures with large variances. This is sensitive to outliers and could obfuscate interesting underlying structures. One of the equivalent definitions of PCA is that it seeks the subspaces that maximize the sum of squared pairwise distances between data projections. This definition opens up more flexibility in the analysis of principal components which is useful in enhancing PCA. In this paper we introduce scales into PCA by maximizing only the sum of pairwise distances between projections for pairs of datapoints with distances within a chosen interval of values [$l,u$]. The resulting principal component decompositions in Multiscale PCA depend on point ($l,u$) on the plane and for each point we define projectors onto principal components. Cluster analysis of these projectors reveals the structures in the data at various scales. Each structure is described by the eigenvectors at the medoid point of the cluster which represent the structure. We also use the distortion of projections as a criterion for choosing an appropriate scale especially for data with outliers. This method was tested on both artificial distribution of data and real data. For data with multiscale structures, the method was able to reveal the different structures of the data and also to reduce the effect of outliers in the principal component analysis.

## 1. Introduction

It is often difficult to extract meaning from multivariate data of high dimension and hence there is a need for feature extraction to make analysis easier and to spot trends, patterns, outliers and other interesting relationship and structures in our data. In 1901, Pearson proposed approximating high dimensional data with lines and planes and hence invented the Principal Component Analysis (PCA). PCA is a linear technique which transforms data to a new coordinate system using linear orthogonal transformation such that the new coordinates are ordered by variance. The coordinate with highest variance is the first principal component; the second principal component is the coordinate with the second highest variance and so on (an example is given in figure 1). PCA is a powerful analysis tool and it is judged to be one of the most important results of applied linear algebra [6] with many interesting applications which include: dimension reduction, blind source separation, data visualization, image compression, and with relevance in many applied disciplines such as quantitative finance, biology, pharmaceutics, taxonomy, healthcare and many more. The principal components from PCA are linear combination of the original components, and even though PCA is limited than non-linear dimension reduction techniques, it is guaranteed to show genuine properties of the original data and the low dimension are meaningful [7].

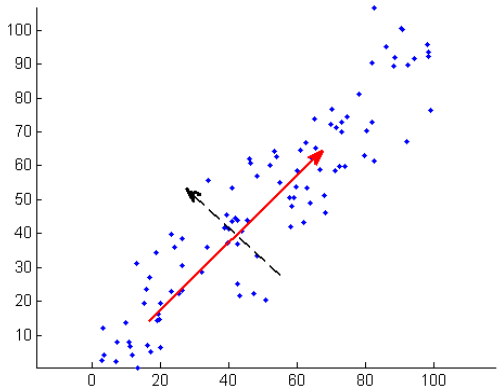

**Figure 1**. Scatter plot of data. The solid red arrow and the dashed black arrow indicate the direction of the first and second principal components respectively. (color online)

However, despite the many applications of PCA, it is not without its drawbacks. An example of such drawbacks is that PCA is based on the covariance matrix which is sensitive to outliers. In this paper, outliers are defined as data elements with large distance from the other data elements in a data sample. Even though outliers can be filtered before performing PCA on the dataset, however in some contexts, identifying outliers could be cumbersome. In addition to the above, datasets are usually noisy (here we define noises as data elements with rather small variance) and the presence of noise in data analysis can further obfuscate the underlying structure(s) of the data being investigated [6].

One of the definitions of PCA is that PCA finds subspaces (lines, planes or higher dimensional subspaces) that maximize the sum of point-to-point squared distances between the orthogonal projections of data points to them.

Let the distance function $dist(x, y)$ be defined by a positive definite quadratic form (Euclidean distance) for pairs of objects **x,y**. In clear terms PCA seeks the $k-$dimensional orthogonal projection that maximizes

$$\sum_{i<j} dist^2(P_L \mathbf{x}_i, P_L \mathbf{x}_j). \tag{1}$$

where $P_L(x)$ is the projection of vector $x$ to plane $L$. We can observe that the maximization problem given above favors large pairwise distances. Hence other interesting structure(s) which can be revealed by smaller pairwise distances may be completely obfuscated. One example of such problem arises when using PCA on data with outliers as the outliers may obfuscate the structure(s) of the data. Let us consider the data shown in figure 2, the data is distributed along a line but with outliers (shown in circles). Figure 3 shows the data projection to the principal components with the *x*-axis as the first principal component. However if the outliers were removed the first principal component should be close to the line on which the data is distributed as shown by the arrow in figure 3.

Figures 4 and 5 are the biplots [1] of the example given above. A biplot is useful for visualizing the magnitude (this is represented with the lines), and sign of each variable's contribution to the first two or three principal components and how each observation (represented as points on the graph) is represented in terms of those components. The axes represent the principal components. From the biplot below, we can observe a significant change in the contribution of the variables in the PCA due to the presence of the outliers.

There are several equivalent definitions of principal components. The definition presented above through maximization of the sum of point-to-point squared distances between the orthogonal projections of data points gives more flexibility for generalization and control [3] which can be manipulated to reveal some interesting underlying structure(s) in our data. In addition to this, the definition above opens up the relationship between PCA and multidimensional scaling [3].

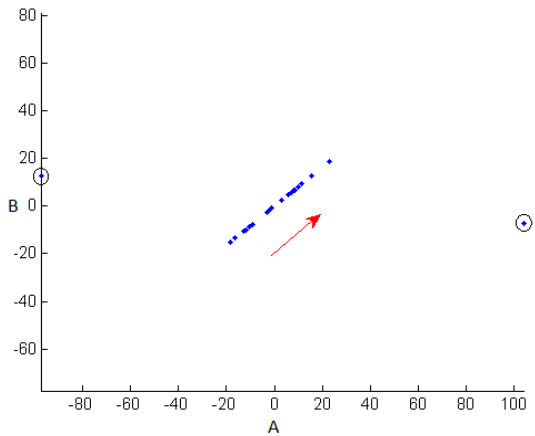

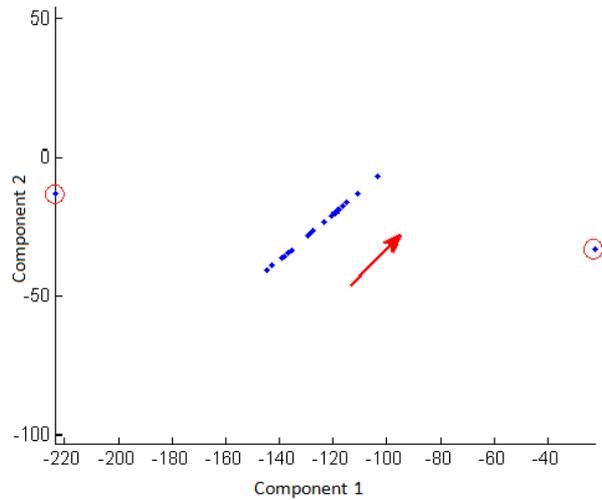

**Figure 2.** Scatter plot of data distributed along a line with 2 outliers. The outliers are shown in circle.

**Figure 3.** Data projection to the first 2 principal components (given by the axes). The arrow indicates the direction of the principal component if the outliers are removed.

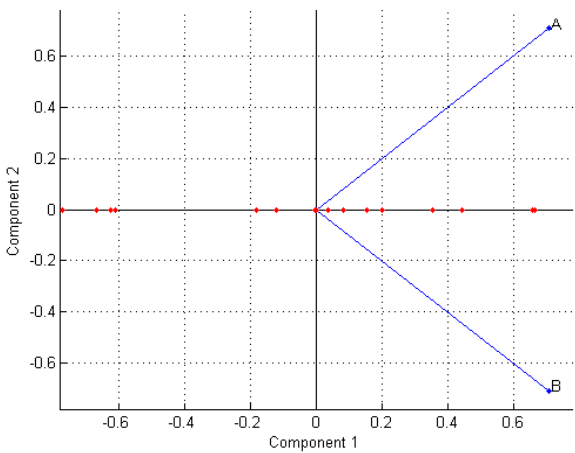

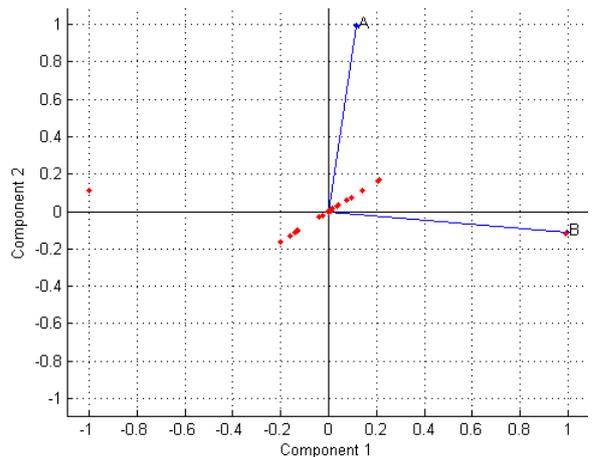

**Figure 4**. The biplot of the data without outliers.

**Figure 5.** The biplot of the data with outliers

In this paper, scale was introduced to enhance the performance of PCA on datasets. That is, we will use in the definition of multiscale PCA maximization of the sum of point-to-point squared distances between the orthogonal projections of data points for the pairs of points with distances in some intervals. The result of this is PCA decomposition of the data which depend on the scale chosen. A further study of these PCA decompositions reveals some underlying structures which could have been obfuscated by other structures such has the presence of outliers or repeated patterns as shown later. We also proposed a criterion for determining the appropriate scale for computing the principal components for data with outliers.

## 2. Definitions and Mathematical Background

In this section we consider four classical approaches to PCA which are equivalent as given by [2] and we also give the necessary mathematical background that will be needed for this paper.

*2.1. Definitions of PCA*

Let $L_k$ be a linear manifold of dimension $k$ *given in the* parametric form as

$L_k = \{\mathbf{v}_0 + a_1\mathbf{v}_1 + a_2\mathbf{v}_2 + ... + a_k\mathbf{v}_k\}$, where $a_i \in \mathbf{R}$, $\mathbf{v}_0 \in \mathbf{R}^m$ and $\{\mathbf{v}_1, \mathbf{v}_2,...,\mathbf{v}_k\}$ is a set of orthonormal vectors in $\mathbf{R}^m$.

Also let $\mathbf{x}_i, i = 1,...,n$ be data elements where $\mathbf{x}_i \in \mathbf{R}^m$ and let the data elements be arranged as the rows of a $n \times m$ matrix $X$ such that the $m$ coordinates is given by the column of $X$. For this paper, the coordinates will be represented by Greek indices while the observations will be represented by Latin indices (i.e. $\mathbf{x}_{i\alpha}$ is the $\alpha$th coordinate of the $i$th observation). For all computations, we assume that the data is centered. This could be achieved by simple translation of the data.

We shall denote the projection of data $\mathbf{x}_i, i=1,2,...,n$ to the plane $L_k$ by $(P_L\mathbf{x}_i)$

*Definition 1 (Data approximation by lines and planes).*
PCA computes the sequences $L_k, (k = 1,2,...,m-1)$ such that the sum of squared distances from data points to their orthogonal projections on $L_k$ is minimal over all linear manifolds of dimension $k$ embedded in $R^m : MSD(X, L_k) \to \min \ (k = 1,2,...,m-1)$.

The mean squared distance between a dataset $X$ and set of vectors $\mathbf{y}$ denoted by $MSD(X, \mathbf{y})$ is defined as $MSD(X, \mathbf{y}) = \sqrt{\frac{1}{n}\sum_{i,j=1}^{N} dist^2(\mathbf{x}_i, P_\mathbf{y}\mathbf{x}_i)}$.

**Remark:** *Dimensionless variables and normalization.* This is exactly the definition given by Pearson in 1901 [5]. Even though Pearson in his paper on principal component analysis did not use in his definition of PCA normalization to unit variance, it is necessary to use the same dimension on all axes. For example we cannot summarise meters with kilograms. Therefore normalization becomes important when the data is from different dimensions; however the choice of normalization should depend on the type of data and the problem being solved.

*Definition 2 (Variance maximization).*
For a dataset $X$ and for a given $\mathbf{v}_i$, let us construct a one-dimensional distribution $\mathbf{B}_i = \{\beta : \beta = \langle \mathbf{x}, \mathbf{v}_i \rangle, \mathbf{x} \in X\}$ where $\langle \cdot, \cdot \rangle$ denotes scalar vector product. Then let us define empirical variance of $X$ along $\mathbf{v}_i$ as $Var(\mathbf{B}^i)$, where $Var()$ is the standard empirical variance. PCA seeks to find such $L_k$ that the sum of empirical variances of $X$ along $\{\mathbf{v}_1, \mathbf{v}_2,...,\mathbf{v}_k\}$ would be maximal over all linear manifolds of dimension $k$ embedded in $R^m : \sum_{i=1,...k} Var(\mathbf{B}_i) \to \max$.

*Definition 3* (*mean point-to-point squared distance maximisation*)

PCA problem consists in finding such sequence $L_k$ that the mean point-to-point squared distance between the orthogonal projections of data points on $L_k$ is maximal over all linear manifolds of dimension $k$ embedded in $\mathbf{R}^m$: $\frac{1}{n}\sum_{i,j=1}^{N} dist^2(P_L\mathbf{x}_i, P_L\mathbf{x}_j) \longrightarrow \max$.

We know that all orthogonal projections onto lower-dimensional space lead to contraction of all point-to-point distances (except for some that do not change), this is equivalent to minimization of mean squared distance distortion: $\sum_{i,j=1}^{N}[dist^2(\mathbf{x}_i, \mathbf{x}_j) - dist^2(P_L\mathbf{x}_i, P_L\mathbf{x}_j)] \longrightarrow \max$.

*Definition 4* (*correlation cancellation*):

PCA seeks such an orthonormal basis $\mathbf{v}_1, \mathbf{v}_2, ..., \mathbf{v}_k$ in which the covariance matrix for $X$ is diagonal. Evidently, in this basis the distributions $\langle \mathbf{v}_i, \mathbf{x} \rangle$ and $\langle \mathbf{v}_j, \mathbf{x} \rangle$, for $i \neq j$, have zero correlation.

*2.2. Mathematics Background*

As earlier stated, PCA seeks the $k$-dimensional projection that maximizes

$$\sum_{i<j} dist^2(P_L\mathbf{x}_i, P_L\mathbf{x}_j). \tag{2}$$

Using the Euclidean distance, this problem can be stated as

$$D_X = \sum_{i<j} \| P_L(\mathbf{x}_i - \mathbf{x}_j) \|_2 \to \max, \tag{3}$$

where $i, j = 1, ..., n$, $L = \sum_{\alpha=1}^{k} a_\alpha v_\alpha$, $a_\alpha \in \mathbf{R}$, $\alpha = 1, 2, ..., k$ and $k \leq m$.

Also $\mathbf{v}_\alpha \in \mathbf{R}^m$ and $\langle \mathbf{v}_\alpha, \mathbf{v}_\beta \rangle = \delta_{\alpha\beta}$ ($\delta$ is kronecker delta).

The projection of a vector $\mathbf{x}$ to a plane $L$ which is denoted by $P_L(\mathbf{x}) = \sum_{\alpha=1}^{k} \mathbf{v}_\alpha (\mathbf{v}_\alpha, \mathbf{x})$.

Therefore the problem (3) reduces to maximizing

$$D_X = \sum_{i<j} \left[ \sum_{\alpha=1}^{k} (\mathbf{v}_\alpha, \mathbf{x}_i - \mathbf{x}_j)^2 \right] \tag{4}$$

This is the same as maximizing the equation (5) below

$$D_X = \sum_{\alpha=1}^{k} \left[ \sum_{i<j} (\mathbf{v}_\alpha, \mathbf{x}_i - \mathbf{x}_j)^2 \right] \tag{5}$$

The expression in the bracket given as

$$\sum_{i<j}(\mathbf{v}_\alpha, \mathbf{x}_i - \mathbf{x}_j)^2 = \sum_{i<j}(\mathbf{v}_\alpha, \mathbf{x}_i - \mathbf{x}_j)(\mathbf{x}_i - \mathbf{x}_j, \mathbf{v}_\alpha)$$
$$= \mathbf{v}_\alpha^T \tilde{S} \mathbf{v}_\alpha. \tag{6}$$

where

$$\tilde{S} = \sum_{i<j}\left[(\mathbf{x}_i - \mathbf{x}_j) \otimes (\mathbf{x}_i - \mathbf{x}_j)\right],$$

and each element of $\tilde{S}$ is given as

$$\tilde{S}_{\alpha\beta} = \sum_{i<j}\left[(\mathbf{x}_{i\alpha} - \mathbf{x}_{j\alpha})(\mathbf{x}_{i\beta} - \mathbf{x}_{j\beta})\right]. \tag{7}$$

Here, $\tilde{S}_{ij}$ is symmetric positive semi-definite because for every $y$, $y \otimes y$ is positive semi-definite. Therefore the problem given by (3) is reduced to

$$\max_{v_1,\ldots,v_k} \sum_{\alpha=1}^{k} \mathbf{v}_\alpha^T \tilde{S} \mathbf{v}_\alpha . \tag{8}$$

Subject to $(\mathbf{v}_\alpha, \mathbf{v}_\beta) = \delta_{\alpha\beta}$ $\alpha, \beta = 1,2,..,k$.

Let $\lambda_1 \geq \ldots \geq \lambda_m$ be the sorted eigenvalues of the matrix $\tilde{S}$ and $\mathbf{e}_1,\ldots,\mathbf{e}_m$ be the corresponding eigenvectors, a maximizer of the constrained maximization problem (8) is $\mathbf{e}_1,\ldots,\mathbf{e}_m$. See theorem 2.1. Hence, the $k$ orthogonal vectors that maximize (8) are the $k$ – eigenvectors corresponding to the highest $k$ – eigenvalues.

If there are $q$ distinct eigenvalues $\lambda_1 \geq \ldots \geq \lambda_q$ of the matrix $\tilde{S}$ such that $\lambda_i$ is of multiplicity $n_i$ and $\sum_{i=1}^{q} n_i = m,$ we have a case called eigenvalue degeneracy. For each $\lambda_i$ with multiplicity $n_i > 1$, the eigenvectors lie in a $n_i$ dimensional subspace orthogonal to the subspace spanned by the non-degenerate eigenvalues. For symmetric matrix, these $n_i$ eigenvectors will be linearly independent and using Gram-Schmidt procedure we can find $n_i$ orthogonal vectors that span this subspace.

Now it is left to show that the solution to the problem (8) is actually the principal components. Let us examine the matrix

$$\tilde{S} = \sum_{i<j}\left[(\mathbf{x}_i - \mathbf{x}_j) \otimes (\mathbf{x}_i - \mathbf{x}_j)\right] = \frac{1}{2}\sum_{i,j=1}^{n}\left[(\mathbf{x}_i - \mathbf{x}_j) \otimes (\mathbf{x}_i - \mathbf{x}_j)\right]$$

$$= \frac{1}{2}\left[n\sum_{i=1}^{n}(\mathbf{x}_i \otimes \mathbf{x}_i) + n\sum_{j=1}^{n}(\mathbf{x}_j \otimes \mathbf{x}_j) - \sum_{i,j=1}^{n}(\mathbf{x}_i \otimes \mathbf{x}_j) - (\mathbf{x}_j \otimes \mathbf{x}_i)\right] \tag{9}$$

$$= \frac{1}{2}\left[2n\sum_{i=1}^{n}(\mathbf{x}_i \otimes \mathbf{x}_i) - n^2(\boldsymbol{\mu} \otimes \boldsymbol{\mu}) - n^2(\boldsymbol{\mu} \otimes \boldsymbol{\mu})\right].$$

Where $\sum_{i,j=1}^{n}(\mathbf{x}_i \otimes \mathbf{x}_j) = \sum_{i=1}^{n}\mathbf{x}_i \otimes \sum_{j=1}^{n}\mathbf{x}_j$ and $\boldsymbol{\mu} = \frac{1}{n}\sum_{i=1}^{n}\mathbf{x}_i$,

$$\tilde{S} = n\sum_{i=1}^{n}(\mathbf{x}_i \otimes \mathbf{x}_i) \tag{10}$$

The remaining terms are zero because the data has been centered

$$\tilde{S} = n^2 \operatorname{cov}(X). \tag{11}$$

We know that given dataset $X$ with empirical covariance matrix $S$, and let $\lambda_k \geq \ldots \geq \lambda_k$ be the sorted eigenvalues of the matrix $S$, the corresponding eigenvector $\mathbf{e}_1,\ldots,\mathbf{e}_k$ is the principal components of the data. From equation (11) above $\tilde{S} = n^2.S$, therefore the eigenvectors of $S$ is also the eigenvectors of $\tilde{S}$ and this is the principal component of $X$ since the multiplication of a matrix by a positive constant does not change the eigenvectors or their order. Hence we have shown that the solution to maximization problem (8) is the principal component of $X$.

Now we consider the elements in the matrix $\tilde{S}$. From equation (7) we have

$$\tilde{S}_{\alpha\beta} = \sum_{i<j} \left[ (\mathbf{x}_{i\alpha} - \mathbf{x}_{j\alpha})(\mathbf{x}_{i\beta} - \mathbf{x}_{j\beta}) \right] \quad (12)$$

$$= \frac{1}{2} \left[ \sum_{i=1}^{n} n \cdot \mathbf{x}_{i\alpha} \mathbf{x}_{i\beta} + \sum_{j=1}^{n} n \cdot \mathbf{x}_{j\alpha} \mathbf{x}_{j\beta} - \sum_{i=1}^{n} \mathbf{x}_{i\alpha} \cdot \sum_{j=1}^{n} \mathbf{x}_{j\beta} \cdot - \sum_{j=1}^{n} \mathbf{x}_{j\alpha} \cdot \sum_{i=1}^{n} \mathbf{x}_{i\beta} \cdot \right]$$

$$= \frac{1}{2} \left[ 2\sum_{i=1}^{n} n \cdot \mathbf{x}_{i\alpha} \mathbf{x}_{i\beta} - \sum_{i=1}^{n} \mathbf{x}_{i\alpha} \cdot \sum_{j=1}^{n} \mathbf{x}_{j\beta} \cdot - \sum_{j=1}^{n} \mathbf{x}_{j\alpha} \cdot \sum_{i=1}^{n} \mathbf{x}_{i\beta} \cdot \right]$$

$$= \sum_{i=1}^{n} n \cdot \mathbf{x}_{i\alpha} \mathbf{x}_{i\beta} - \sum_{i=1}^{n} \mathbf{x}_{i\alpha} \cdot \sum_{j=1}^{n} \mathbf{x}_{j\beta}$$

$$= \sum_{i,j=1}^{n} (n \cdot \delta_{ij} - 1) \mathbf{x}_{i\alpha} \mathbf{x}_{j\beta} = \sum_{i,j=1}^{n} L_{ij} \mathbf{x}_{i\alpha} \mathbf{x}_{j\beta}. \quad (13)$$

In matrix notation, the quadratic form (13), can be written as

$$\tilde{S}_{\alpha\beta} = (X_{\alpha}^{T} L X_{\beta})$$
$$\tilde{S} = (X^{T} L X). \quad (14)$$

Where $L = [L_{ij}] = n \cdot \delta_{ij} - 1$ and $\delta_{ij}$ is the kronecker delta. $L$ is an $n \times n$ symmetric positive-semi definite matrix with zero column and row sum and this is useful for describing the pairwise relationship between data elements as shown in Lemma 2.1 which is also available in [7].

**Theorem 2.1:** Let A be an $n \times n$ symmetric matrix and let the sorted eigenvalues be given by $\lambda_1 \geq \lambda_2 \geq \ldots \geq \lambda_n$ and let $e_1, \ldots, e_n$ be the corresponding eigenvectors. Then $e_1, \ldots, e_n$ is a maximizer of the constrained maximization problem

$$\max_{\mathbf{u}_1, \ldots, \mathbf{u}_k} \sum_{\alpha=1}^{k} \mathbf{u}_{\alpha}^{T} A \mathbf{u}_{\alpha}$$

Subject to: $(\mathbf{u}_{\alpha}, \mathbf{u}_{\beta}) = \delta_{\alpha\beta}, \alpha, \beta = 1, 2, \ldots, k$.

A detailed proof of this theorem is available in [7]

**Lemma 2.1:** Let $L$ be as defined above and let $\mathbf{x} \in \mathbf{R}^n$ then

$$\mathbf{x}^T L \mathbf{x} = \sum_{i<j}^{n} -L_{ij} (\mathbf{x}_i - \mathbf{x}_j)^2.$$

And for k coordinate vectors we have:

$$\sum_{\alpha=1}^{k} \mathbf{x}_{\alpha}^{T} L \mathbf{x}_{\alpha} = \sum_{i<j}^{n} -L_{ij} (\sum_{\alpha=1}^{k} (\mathbf{x}_{i\alpha} - \mathbf{x}_{j\alpha})^2$$

$$= \sum_{i<j}^{n} -L_{ij} \| \mathbf{x}_i - \mathbf{x}_j \|^2.$$

Hence we see that the matrix given by $L$ is useful because the quadratic form associated with it is the weighted sum of all pairwise squared distances [7].

*2.3. Weighted PCA*

Definition 3 allows for some flexibility in the analysis of principal components because we have control over the pairwise distances of projected data. By assigning weights to these pairwise distances, we can manipulate the resulting PCA decomposition of the data.

We now consider the problem of finding the principal component using weighted pairwise distances of projected data. This problem is stated below.

$$\sum_{i<j} w_{ij}[dist^2(P_L\mathbf{x}_i, P_L\mathbf{x}_j)]$$

$$D_X = \sum_{i<j} w_{ij} \| P_L(\mathbf{x}_i - \mathbf{x}_j)\|^2 \to \max \qquad (15)$$

Subject to: $(\mathbf{v}_\alpha, \mathbf{v}_\beta) = \delta_{\alpha\beta}.$

Where $w_{ij} = w_{ji}$ is the non-negative weight assigned to the distance between element $i$ and $j$ and $w_{ij} = 0,$ for $i = j$.

The equation (15) reduces to maximizing the equation 16 below

$$D_X = \sum_{i<j} w_{ij} \left[ \sum_{\alpha=1}^{k} (\mathbf{v}_\alpha, \mathbf{x}_i - \mathbf{x}_j)^2 \right]. \qquad (16)$$

This is the same as

$$D_X = \sum_{\alpha=1}^{k} \left[ \sum_{i<j} w_{ij} (\mathbf{v}_\alpha, \mathbf{x}_i - \mathbf{x}_j)^2 \right].$$

The expression in the bracket given as

$$\sum_{i<j} w_{ij} (\mathbf{v}_\alpha, \mathbf{x}_i - \mathbf{x}_j)^2 = \sum_{i<j} w_{ij} (\mathbf{v}_\alpha, \mathbf{x}_i - \mathbf{x}_j)(\mathbf{x}_i - \mathbf{x}_j, \mathbf{v}_\alpha)$$

$$= \mathbf{v}_\alpha^T \tilde{M} \mathbf{v}_\alpha, \qquad (17)$$

where

$$\tilde{M} = \sum_{i<j} w_{ij} \left[ (\mathbf{x}_i - \mathbf{x}_j) \otimes (\mathbf{x}_i - \mathbf{x}_j) \right] \qquad (18)$$

Let $R_i = \sum_j w_{ij}$ and let $C_j = \sum_i w_{ij}$.

Equation (18) can be written as

$$\tilde{M} = \sum_{i<j} w_{ij} \left[ (\mathbf{x}_i - \mathbf{x}_j) \otimes (\mathbf{x}_i - \mathbf{x}_j) \right] = \frac{1}{2} \sum_{i,j=1}^{n} w_{ij} \left[ (\mathbf{x}_i - \mathbf{x}_j) \otimes (\mathbf{x}_i - \mathbf{x}_j) \right] \qquad (19)$$

$$= \frac{1}{2}\left[\sum_{i=1}^{n} R_i(\mathbf{x}_i \otimes \mathbf{x}_i) + \sum_{j=1}^{n} C_j(\mathbf{x}_j \otimes \mathbf{x}_j) - \sum_{i,j=1}^{n} w_{ij}[(\mathbf{x}_i \otimes \mathbf{x}_j) + (\mathbf{x}_j \otimes \mathbf{x}_i)]\right] \quad (20)$$

$$= \frac{1}{2}\left[\sum_{i=1}^{n} (R_i + C_i)(\mathbf{x}_i \otimes \mathbf{x}_i) - \sum_{i,j=1}^{n} w_{ij}[(\mathbf{x}_i \otimes \mathbf{x}_j) + (\mathbf{x}_j \otimes \mathbf{x}_i)]\right]. \quad (21)$$

Each element is given as

$$M_{\alpha\beta} = \frac{1}{2}\left[\sum_{i=1}^{n} (R_i + C_i)\mathbf{x}_{i\alpha}\mathbf{x}_{i\beta} - \sum_{i,j=1}^{n} w_{ij}(\mathbf{x}_{i\alpha}\mathbf{x}_{j\beta} + \mathbf{x}_{j\beta}\mathbf{x}_{i\alpha})\right] \quad (22)$$

$$M_{\alpha\beta} = \left[\sum_{i=1}^{n} R_i(\mathbf{x}_{i\alpha}\mathbf{x}_{i\beta}) - \sum_{i,j=1}^{n} w_{ij}(\mathbf{x}_{i\alpha}\mathbf{x}_{j\beta})\right] \quad (23)$$

because $R_i = C_i$ and $\sum_{i,j=1}^{n} w_{ij}(\mathbf{x}_{i\alpha}\mathbf{x}_{j\beta}) = \sum_{i,j=1}^{n} w_{ij}\mathbf{x}_{j\beta}\mathbf{x}_{i\alpha}$.

Therefore, we can write equation (23) as

$$M_{\alpha\beta} = (\delta_{ij} R_i - w_{ij})\mathbf{x}_{i\alpha}\mathbf{x}_{j\beta},$$

$$M_{\alpha\beta} = \left(\delta_{ij}(\sum_{j=1}^{n} w_{ij}) - w_{ij}\right)\mathbf{x}_{i\alpha}\mathbf{x}_{j\beta}. \quad (24)$$

Let $L^w = [L^w{}_{ij}] = \left(\delta_{ij}(\sum_{j=1}^{n} w_{ij}) - w_{ij}\right)$. This can be written in the form below.

$$L_{ij} = \begin{cases} \sum_{j=1}^{n} w_{ij} & i = j \\ -w_{ij} & i \neq j \end{cases}$$

Where $w_{ij} = 0$, for $i = j$.

In matrix notation, the quadratic form (24), can be written as

$$\tilde{M}_{\alpha\beta} = (X_\alpha^T L^w X_\beta) \quad (25)$$

and
$$\tilde{M} = (X^T L^w X).$$

Therefore the problem given by (15) is reduced to

$$\max_{v_1,\ldots,v_k} \sum_{\alpha=1}^{k} \mathbf{v}_\alpha^T \tilde{M} \mathbf{v}_\alpha$$

Subject to $(\mathbf{v}_\alpha, \mathbf{v}_\beta) = \delta_{\alpha\beta}$  $\alpha, \beta = 1,2,..,k$, \quad (26)

where $\tilde{M}$ is a symmetric positive semi-definite matrix, and from theorem 2.1, the eigenvectors corresponding to the sorted eigenvalues of the matrix $\tilde{M}$ is a maximizer of the constrained maximization problem (26). In the case of degenerated eigenvalues, the set $\mathbf{e}_1,\ldots,\mathbf{e}_m$ is not uniquely defined.

## 3. Multiscale PCA (MPCA)

In this section, we introduce the Multiscale PCA (MPCA) algorithm to enhance the robustness of the PCA especially in revealing hidden structure(s) that may be present in dataset but which the conventional approach might not reveal. MPCA compute principal components by maximizing the sum of pairwise distances between data projection for only pairs of datapoints for which the distance is within the chosen scale. This is achieved by assigning a weight of 1 to the pairwise distance of projections of any pair of data points with distance within the chosen scale and a weight of 0 otherwise.

$$\begin{cases} w_{ij} = 1 & l \leq \| \mathbf{x}_i - \mathbf{x}_j \|_2 \leq u \\ w_{ij} = 0 & otherwise. \end{cases} \quad (27)$$

In the scale interval, $(l,u)$, $l$ is the lower limit of the scale and $u$ is the upper limit. Let $d^{min}$ be the minimum pairwise distance greater than zero and $d^{max}$ be the maximum pairwise distance in the data. We select the pairs $(l,u)$ from a triangle $\Delta = \{(l,u): d^{min} \leq l < u \leq d^{max} \}$.

With this control over the pairwise distances, we are able compute PCA at various scales and the outcome of this is scale dependent PCA which can reveal interesting underlying structure(s) that may be present in data. For example, reducing the upper limit of the scale while keeping the lower limit at 0 translate to computing PCA by considering smaller distances and excluding very large distances. This has the effect of minimizing without explicit exclusion the contribution of certain influential data elements in the analysis of the principal components.

*3.1. The Multiscale PCA Algorithm*

Here we discuss the Multiscale PCA Algorithm.

1. Given the data sample.
2. Centralize the data by subtracting the mean of the variables from each observation.
3. Find the dissimilarity matrix by computing the Euclidean distance.
4. Choose an appropriate scale between 0 and the maximum distance. For easy analysis, a scale between 0 and 1 could be chosen and then multiplied by the maximum distance. For this paper when using scale between 0 and 1 we call it standard scale.
5. Calculated the binary weight as given in equation (27)
6. Calculate the matrix $L^w$ as given below

$$L_{ij}^w = \begin{cases} \sum_{j=1}^{n} w_{ij} & i = j \\ -w_{ij} & i \neq j \end{cases}$$

7. Calculate the matrix $A = Y^T L^w Y$, where $Y$ is the centralized data.
8. Find the sorted eigenvalues of the matrix $A$ in descending order of magnitude and project the data onto their corresponding eigenvectors. This will be the principal components at the selected scale.

To illustrate the result of MPCA on data, we consider some examples.

*3.2. Multiscale PCA on Data with repeated patterns*

**Example 1**

Here we consider an example of a data sample with repeated underlying structure. See figure 6-9.

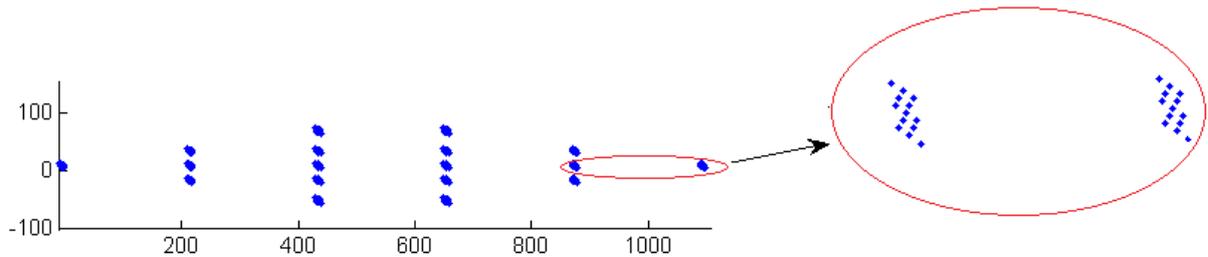

**Figure 6.** Scatter plot of data with repeated pattern.

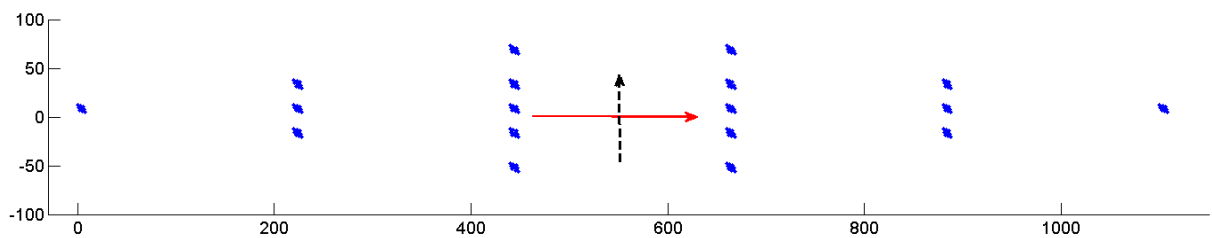

**Figure 7.** The solid arrow and the dashed arrows show the direction of the first and second principal components respectively using MPCA at a scale of [0-1108] equivalent to standard scale [0-1]. This is the same as using PCA .

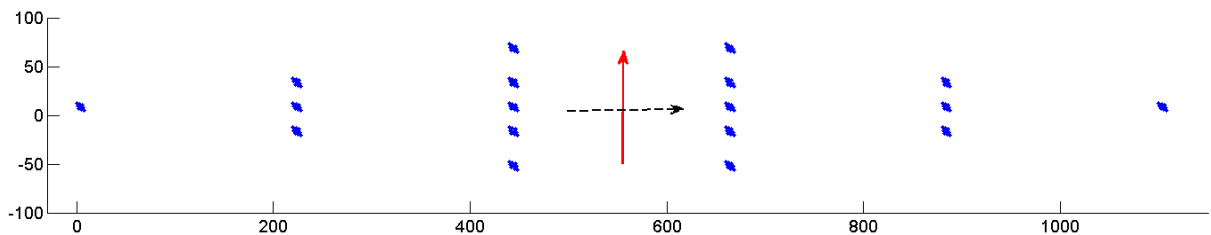

**Figure 8.** The solid arrow and the dashed arrow show the direction of the first and second principal components respectively using MPCA at a scale of [0-200] equivalent to standard scale [0-0.18].

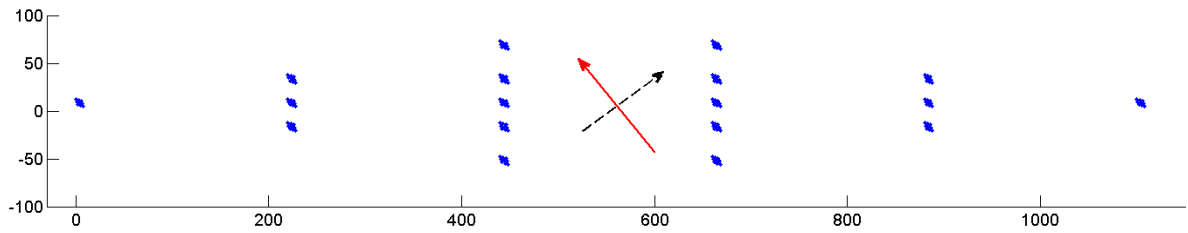

**Figure 9.** The solid arrow and the dashed arrow show the direction of the first and second principal components respectively using MPCA at a scale of [0-12] equivalent to standard scale [0-0.01].

From figure 9 we observe that the PCA reveals the inner structure of the data. A better view of this inner structure and the PCA is given in figure 10.

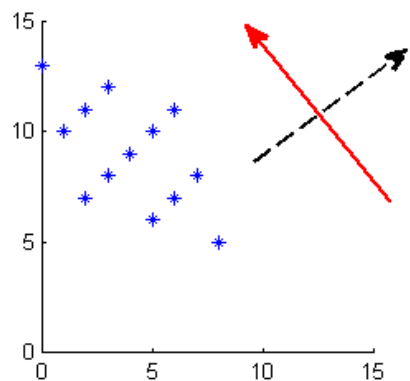

**Figure 10.** Magnified view of one cluster in the dataset with the solid red arrow and dashed black arrow representing the direction of the first and second principal components respectively (Color online).

As illustrated in the example above, the principal components changed as the scale changed and this was able to reveal some underlying structures of the data. Figure one captures the changes in the first principal component at various scales.

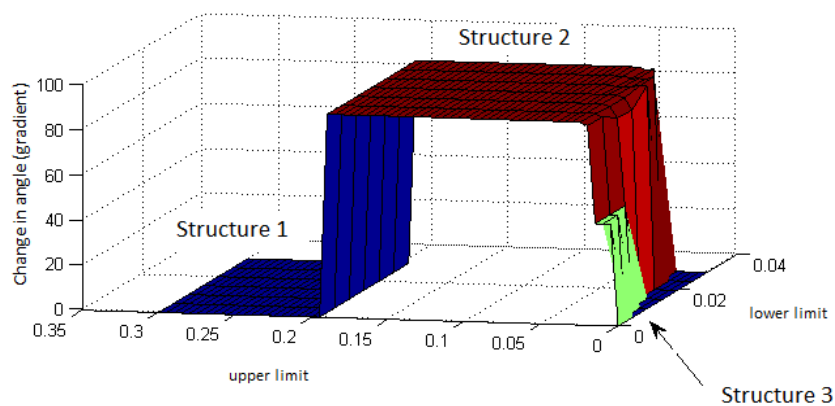

**Figure 11.** The diagram illustrate the change in the angle of the first principal component as the scale changed. The angle recorded here is the angle (in gradient) between the first principal component using PCA and the first principal component using MPCA at a given scale.

## 4. Clustering Analysis on the Interval of Scales

To further study these structures, we consider clustering analysis on the interval of scales and we introduce the Ratio of Distortion in this section.

### 4.1. Representing PCA Structures.

Let us consider the interval of values $[l,u]$ where $l =$ lower limit, $u =$ upper limit and $l < u$. The scale $(l,u)$ can be represented as point in the plane $\mathbf{R}^2$ as shown in figure 12. The resulting principal component decompositions in MPCA depend on the points $(l,u)$ on the plane.

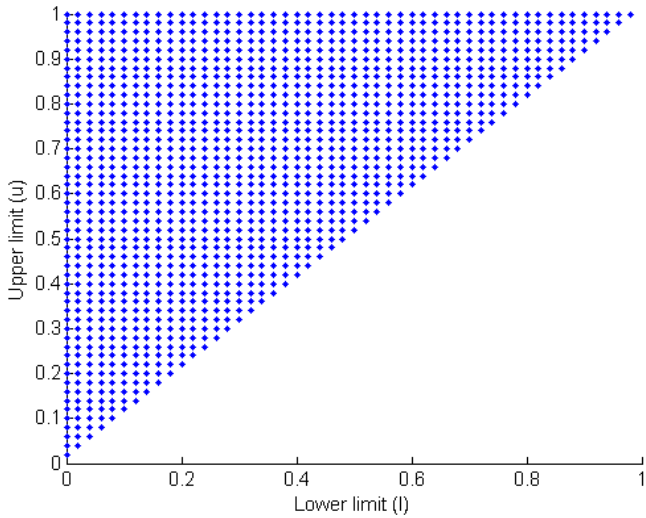

**Figure 12.** This diagram shows the standard scale represented as points on the plane $\mathbf{R}^2$.

We will like to study the PCA structure at different scales; therefore we need a representation of the PCA structure for each point $(l,u)$ on the plane. We can represent the PCA structure at a point $(l,u)$ by the corresponding orthonormal vectors of principal components from MPCA at that point; however this representation is not convenient for statistical analysis of principal components.

If for example we consider the case of equidistribution of a normalized vector $\mathbf{v}$ on $m-1$ sphere the expectation $E[\mathbf{v}] = 0$. This is because of spherical symmetry and the Expectation is the vector in the sphere which is rotation invariant and that is $0$, and this could be counter intuitive. The space of principal component bases is a space of orthonormal bases in $R^m$. This is not a linear space but a rather complicated symmetric manifold with group $O_m$ action on it. We propose to embed this symmetric space into a Euclidean space using the PCA projector representation and, after that, apply standard statistical and data mining procedures. Let us recall that the principal component given by $\mathbf{e}_i$ is the same as $-\mathbf{e}_i$, therefore we need a representation such that this condition is satisfied. The principal components are orthogonal axial frame [7] and one way to represent such data is using the tensor product $P_i = \mathbf{e}_i \otimes \mathbf{e}_i$, which is the projector of our data onto the principal component $\mathbf{e}_i$.

Since this product is bi-linear we know that $-\mathbf{e}_i \otimes -\mathbf{e}_i = \mathbf{e}_i \otimes \mathbf{e}_i$, hence we have the same representation for both a vector and its negative as required.
$P_i X = \mathbf{e}_i (\mathbf{e}_i, X) = (\mathbf{e}_i \otimes \mathbf{e}_i) X$ is the projection of data $X$ onto vectors $\mathbf{e}_i$ and

$$\sum_{i=1}^{k} P_i X = \sum_{i=1}^{k} \mathbf{e}_i (\mathbf{e}_i, X)$$ is the data $X$ projected onto the first $k$ - principal component.

For any $m$ orthonormal vectors $\mathbf{e}_1,\ldots,\mathbf{e}_m$, $\sum_{i=1}^{m}\mathbf{e}_i \otimes \mathbf{e}_i = 1$. If $\mathbf{e}$ is one of $\mathbf{e}_i$ with probability $\frac{1}{m}$, then $E(\mathbf{e}\otimes\mathbf{e}) = \frac{1}{m}$. The rotation invariance gives the same result if $\mathbf{e}$ is equidistributed on unit $m-1$ sphere.

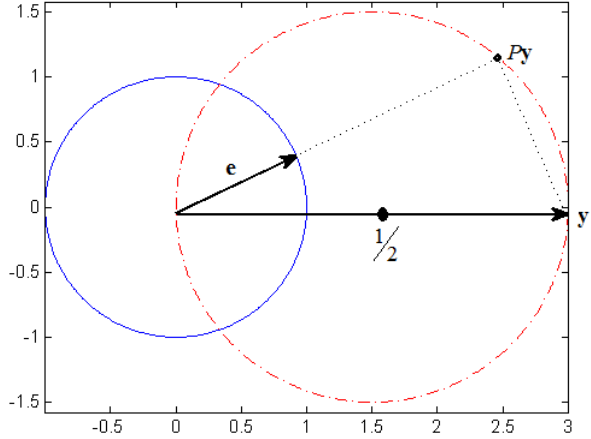

**Figure 13.** This diagram illustrates how the projection of vector $\mathbf{y}$ changes (given by the dashed red sphere) as $\mathbf{e}$ moves along the blue solid 1-sphere (Color online).

It follows that the average projection of $E[\sum_{i=1}^{k} P_i X] = \frac{k}{m} X$.

Therefore we represent the PCA structure of the data at any point $(l,u)$ by the sum of the projectors corresponding to MPCA at that point. This will be denoted by $\rho_k = \sum_{i=1}^{k}\mathbf{e}_i \otimes \mathbf{e}_i$, $k = 1,2,\ldots,m-1$.

The full description of the principal components decompositions of data $X$ is given by an ordered set ("cortege") of matrices $\rho_1, \rho_2,\ldots,\rho_{m-1}, \rho_m = 1$.

If we arrange the $\mathbf{e}_i$ $i = 1,2,\ldots,k$ as columns of matrix $E$, then

$\rho_k X = EE^T X$ and $\rho_k = EE^T$. For $k = m$, $EE^T = \mathbf{I}$.

MPCA lead to scale dependent PCA structures and with these PCA structures represented as defined above, we can study the structures in our data further by analyzing these projectors.

The PCA structures associated with two different points on the plane is said to be similar if their corresponding projectors $\rho_k$ are similar

*4.2. Clustering of Scales.*
We guess that in some cases there are clear internal structures in the data which depend on scales. Performing MPCA on the data leads to a continuum of PCA structures depending on scales used and to reveal the structures in the data, we join scales with similar PCA structures and separate scales with dissimilar PCA structures. This leads to the idea of clustering of scales.

We represent the distance between two points on the scale by the distance between their corresponding PCA structures. Clustering analysis of the scales group similar PCA structures together and this reveals some structures in the data. We describe each cluster by the projector corresponding to the medoid point of the cluster. In a later section, we will introduce Ratio of Distortion which is another criterion that can be used to select the projectors that describe the clusters.

For example, clustering analyses of scales for $\rho_2$ corresponds to cluster analysis of the MPCA structures when data is projected onto the first 2 principal components at various scales.

Now let each point $(l,u)$ in the plane be represented by $\chi_p$, where $\chi_p = (l,u)$, $l \in L$, $u \in U$ such that $l < u$. We denote the projector $\rho_k$ at a point $\chi_p$ by $\rho_{\chi_p}$. For any pair of points $\chi_p, \chi_q$ in the space of scales we can compute the distance between the associated projectors $\rho_{\chi_p}, \rho_{\chi_q}$ for a given $k$ using invariant norm. We recall that the Frobenius norm of a real matrix $B$ denoted by $\|B_F\| = \sqrt{trace\{BB^T\}}$, therefore distance between projectors of any pair of points in the space of scale
$$dist^2(P_{\chi_p}, P_{\chi_q}) = \sqrt{trace\{(P_{\chi_p} - P_{\chi_q})^T (P_{\chi_p} - P_{\chi_q})\}}.$$

Any standard clustering algorithm can be used to cluster the scale in order to reveal hidden structures in the data but in this paper, agglomerative hierarchical clustering was used because we can measure distance easily. Deciding on the number of true clusters in clustering analysis is a classical problem and one may want to compare various indices. A typical example of such is the $pseudo\ t^2$ statistic.

$$pseudo\ t^2 = \frac{[SSE_t - (SSE_a + SSE_b)](n_a + n_b - 2)}{SSE_a + SSE_b}.$$

Where $SSE_a$ is the sum of square of cluster $a$, $SSE_b$ is the sum of square of cluster $b$, $SSE_t$ is the sum of square of cluster formed by joining clusters $a$ and $b$, $n_a$ and $n_b$ are the number of elements in clusters $a$ and $b$ respectively. If a small value of the $pseudo\ t^2$ statistic at a step $i$ of the hierarchical clustering is followed by a distinct large value at the step $i+1$, the cluster form at the step $i$ is chosen as the optimal cluster. It is assume that the mean vector of the two clusters being merged at the step $i+1$ can be regarded as different and should probably not be merged.

Let us consider the result of the cluster analysis of the data in example 1 (figure 6) for $\rho_1$ (i.e. projection onto first principal component). For illustration purpose points from the subset of $L$ and $U$ have been selected. $l \in \{(0, 0.4), (0.1, 0.95)\}$ and $u \in \{(0.005, 0.01), (0.11, 0.19), (0.2, 1)\}\ u$. The result is presented in figure 14.

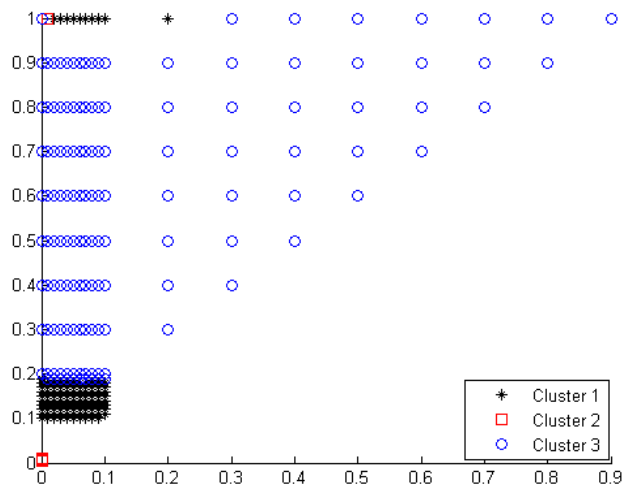

**Figure 14.** This diagram shows cluster of scales on the plane. Scales belonging to the same cluster are represented by the same symbol and color. (Color online)

The $pseudo\ t^2$ statistic indicates three meaningful clusters. This reaffirms the result displayed in figure 11.

We represent each of these structures by the eigenvector of the medoid point of the cluster representing it. The result is given in the table 1.

**Table 1.** This table shows the description of each cluster. Each cluster has been described by the eigenvector of the projector corresponding to the medoid point of the cluster.

| Cluster | Interval corresponding to Medoid point (Projector) | Eigenvector |
|---|---|---|
| 1 | (0.1,0.19) | $\mathbf{e}_1 = [-0.0019,\ 1.0000]$ |
| 2 | (0,0.01) | $\mathbf{e}_1 = [-0.7071\ \ 0.7071]$ |
| 3 | (0.3,0.8) | $\mathbf{e}_1 = [-1.0000,\ 0.0000]$ |

**Example 2**

We consider MPCA of the 'Energy Efficiency Dataset' available online at the UCI machine Learning Repository. This dataset contains 768 samples and 8 components and used to predict 2 different outputs. We perform MPCA on the data (the output variables are not included), since all the data are positive, we normalized by dividing by the mean. The data projections to the first two principal component and first three principal components respectively are shown in the figures 15 and 16. MPCA at standard scale of [0-0.2] reveals the structure given in figures 17 and 18.

The result of the clustering analysis of the scale is presented in figure 19. The $pseudo\ t^2$ statistic indicates four meaningful clusters.

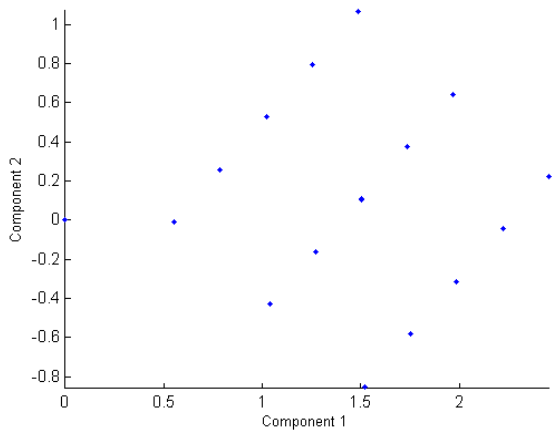

**Figure 15.** Data projection to the first 2 principal components for PCA.

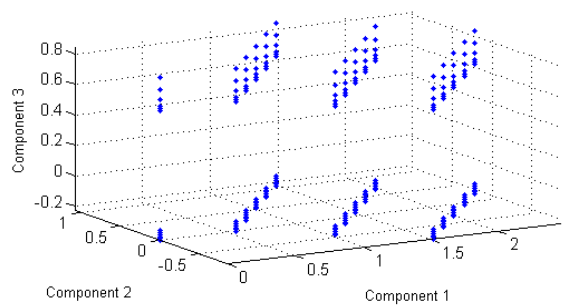

**Figure 16.** Data projection to the first 3 principal components for PCA.

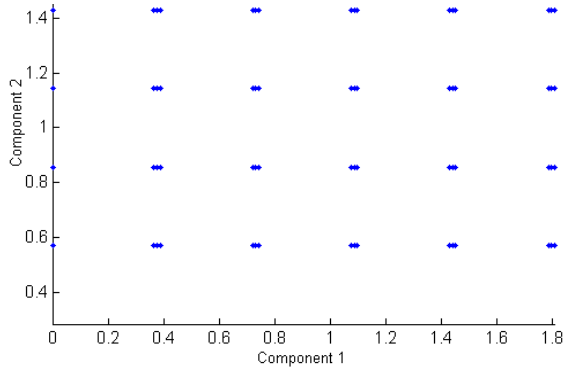

**Figure 17.** Data projection to the first 2 principal components using MPCA at standard scale (0, 0.2).

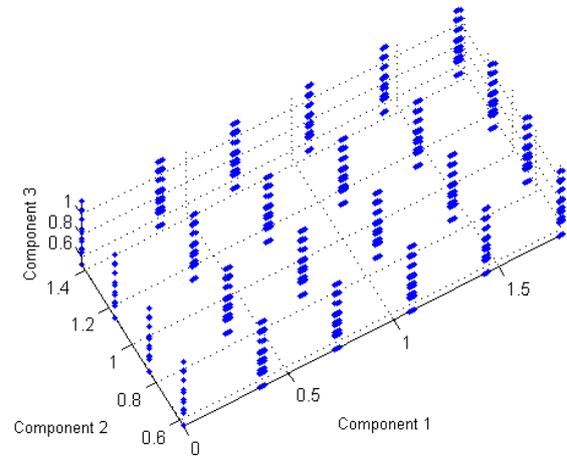

**Figure 18.** Data projection to the first 3 principal components using MPCA at standard scale (0, 0.2)

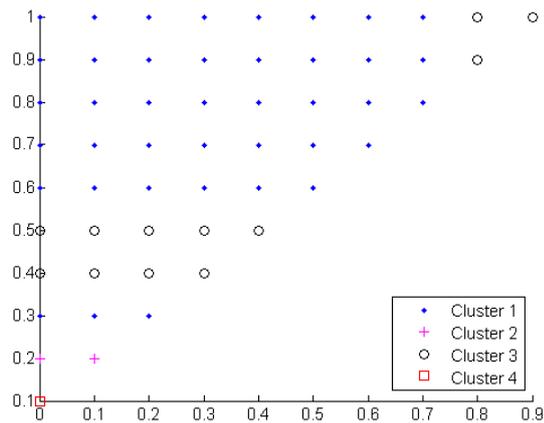

**Figure 19.** This diagram shows cluster of scales on the plane. Scales belonging to the same cluster are represented by the same symbol and color. (Color online)

We represent each of these structures by the eigenvector of the medoid point of the cluster representing it. The result is given in the table 2.

**Table 2.** This table shows the description of each cluster. Each cluster has been described by the eigenvectors of the projector corresponding to the medoid point of the cluster.

| Cluster | Interval corresponding to Medoid point | Eigenvector |
|---|---|---|
| 1 | (0,0.9) | $\mathbf{e}_1 = [0, 0, 0, 0, 0, 0, -0.7618, -0.6478]$ |
|   |   | $\mathbf{e}_2 = [0, 0, 0, 0, 0, 0, -0.6478, -0.7618]$ |
| 2 | (0,0.2) | $\mathbf{e}_1 = [0, 0, 0, 0, 0, 0, -0.0172, 0.9999]$ |
|   |   | $\mathbf{e}_2 = [0, 0, 0, 0, 0, 1.0000, 0, 0]$ |
| 3 | (0.9-1) | $\mathbf{e}_1 = [0, 0, 0, 0, 0, 0, -0.6950, -0.7190]$ |
|   |   | $\mathbf{e}_2 = [0.2664, -0.2587, 0.0770, -0.5614, 0.7355, 0, 0, 0]$ |
| 4 | 0-0.1 | $\mathbf{e}_1 = [0.4266, -0.4185, -0.7982, -0.0762, 0, 0, 0]$ |
|   |   | $\mathbf{e}_2 = [-0.3860, 0.2288, -0.4024, -0.7979, 0, 0, 0, 0]$ |

*4.3. Multiscale PCA on Data with Outliers*

The presence of outliers in our data serves to obfuscate the underlying structure of the data in PCA. MPCA is however effective in revealing the underlying structure of data with outliers. By reducing the upper limit of the scale, we can effectively mitigate the effect of outliers in the analysis of the principal components without explicit exclusion of these outliers.

**Example 3**

To test the performance of scaled PCA on data with outliers, data were simulated along known plane and some outliers were added to this data. This data was embedded into a higher dimensional space and we seek to recover the original plane from the data by using PCA and MPCA (at various scales). The angle between the original directional vector and the first principal component of MPCA at various scales is given in the appendix (see table A1).

We consider a 3-dimensional data sample in which the elements are distributed uniformly on a plane (2-d) with the directional vectors given as

$\mathbf{u} = [0.8944, -0.4472, 0.0000]$;

$\mathbf{v} = [0.1826, 0.3651, -0.9129]$;

With vector $\mathbf{u}$ being the first principal component and few outliers were added as can be seen in figure 19. The projection of the data to the first 2 principal components is shown in figure 20; this has been influenced by the outliers in the data. MPCA at standard scale of (0-0.8) however gives another structure which is found to have captured the data quite well as shown in figure 21. The result of the clustering analysis of the scales is presented in figure 22.

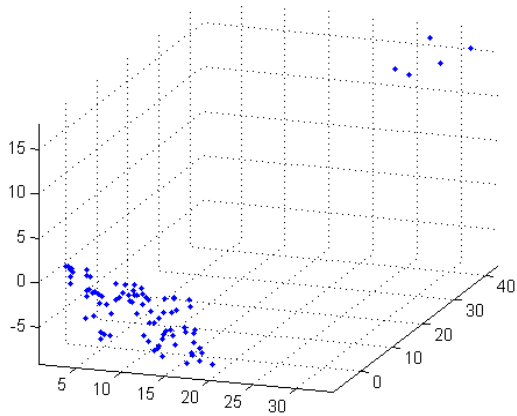

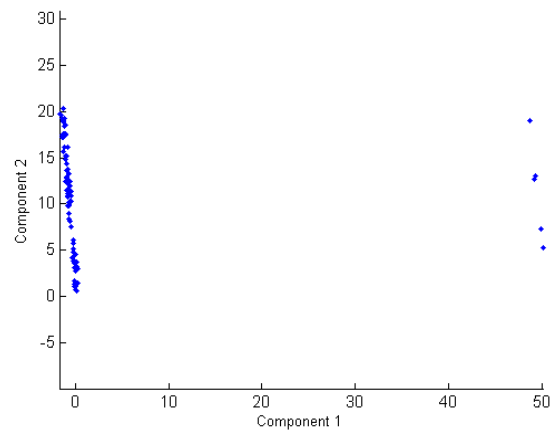

**Figure 19.** Scatter plot of data in 3-dimension with a few outlying points.

**Figure 20.** Data projection to the first 2 principal components using PCA. It can be observed that the outliers have influenced the result of the PCA.

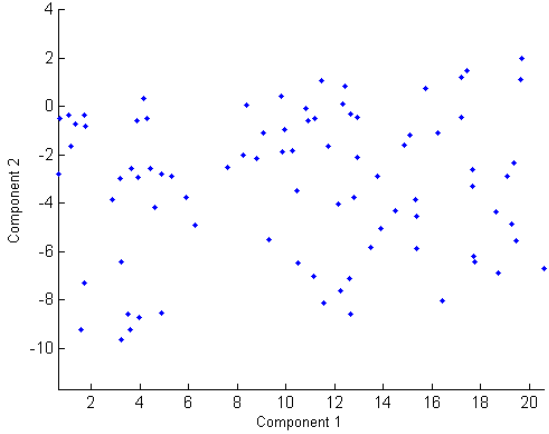 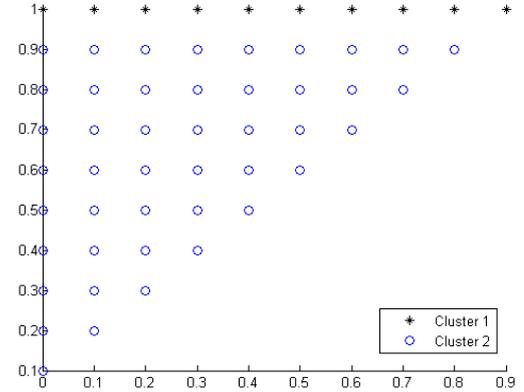

**Figure 21.** Data projection to the first 2 principal components using MPCA at standard scale of (0, 0.8). The effects of the outliers have been mitigated.

**Figure 22.** This diagram shows cluster of scales on the plane. Scales belonging to the same cluster are represented by the same symbol and color (Color online).

We represent each of these structures by the eigenvector of the medoid point of the cluster representing it. See table 3.

**Table 3.** This table shows the description of each cluster, described by the eigenvectors of the projector ($\rho_2$) corresponding to the medoid point of the cluster.

| Cluster | Interval corresponding to Medoid point (Projector) | Eigenvector |
|---|---|---|
| 1 | (0.4,1) | $e_1 = [-0.3561, -0.8579, -0.3704]$<br>$e_2 = [-0.8613, 0.4551, -0.2261]$ |
| 2 | (0,0.1) | $e_1 = [-0.9113, 0.3382, 0.2349]$<br>$e_2 = [0.0538, -0.4679, 0.8821]$ |

*4.4. Criterion for choosing scale for data with outliers*

In this section we propose a criterion for deciding an appropriate scale for MPCA in feature extraction especially for data with outliers.

As mentioned earlier, finding the principal components using definition 3 is equivalent to minimizing the mean squared distance distortion.

$$\sum_{i,j=1}^{N}[dist^2(\mathbf{x}_i,\mathbf{x}_j) - dist^2(P_L\mathbf{x}_i,P_L\mathbf{x}_j)] \longrightarrow \min$$

Where the dimension $k$ of $L$ is strictly less than the dimension of the data.

Hence we propose that an appropriate scale for a given dimension $k$ could be determined by finding the ratio of distortion

$$\frac{\sum_{i,j=1}^{N} \|(P_L \mathbf{x}_i) - (P_L \mathbf{x}_j)\|^2}{\sum_{i,j=1}^{N} \|(\mathbf{x}_i - \mathbf{x}_j)\|^2} \longrightarrow \max,$$

For all $\mathbf{x}_i, \mathbf{x}_j$ such that $l \leq \|\mathbf{x}_i - \mathbf{x}_j\|^2 \leq u$

$l$ Is the lower limit of the scale and $u$ is the upper limit.

The ratio of distortion introduced here can also be used in the clustering analysis of scales as a criterion to determine the PCA structure that describes the cluster.

## 5. Discussion and Conclusion

### 5.1. Discussion

For example 3, MPCA at scales $(l,u)$, $0 \leq l \leq 0.3$, $0 < u \leq 0.9$, $l < u$, reveals another structure in the data that has been obfuscated by the outliers. Table A2 and A3 in the appendix show the results of the ratio of distortion for 2 different dimensions. This is also consistent with the difference in angle between the original plane and the principal component computed using MPCA at these scales (see table A1 in the appendix).

Reducing the upper limit to a very small number may cause MPCA to fit noise while increasing the lower limit only may cause MPCA to fit outliers if such is present in the data. It is important to note that by using MPCA, some pairwise distances are exempted in the analysis of principal component and the percentage of such exempted pairwise distance should be kept to a reasonable number.

As it can be observed in the table A2 and A3 in the appendix, as the lower limit increased, the ratio of distortion appear to improve (even though the difference in angle is quite large for some scales) but only because MPCA is fitting outliers. Therefore, in addition to the result of the ratio of distortion, the percentage of total pairwise distances exempted in the computation of the MPCA at different scales (especially when $l > 0$) should be considered in choosing an appropriate scale. A good scale for MPCA should be one with maximum ratio of distortion and least number of exempted pairwise distances.

Table A4 in the appendix shows the percentage of pairwise distances of data points exempted in the computation of MPCA at various scales. It can be concluded that while reducing the upper limit is good for this data, increasing the lower limit makes MPCA to fit outliers.

### 5.2. Conclusion

Principal component analysis of high dimension data favour components with high variance. This may obfuscate hidden geometric structures that may be present in the data. The definition of PCA as the maximization of the sum of point-to-point squared distances between the orthogonal projections of data points is a very convenient definition and allows for generalization. In this paper, we introduced multiscale PCA as maximization of the sum of point-to-point squared distances between the orthogonal projections of data points for the pairs of points with distances in some intervals (scales). MPCA is developed to solve the problem of revealing hidden geometric structures in data. The result of MPCA on data leads to a continuum of PCA structures of the data which is dependent on the intervals chosen. Analysing the MPCA structures of data reveals some internal structures of the data especially for data with multiscale structures. To study the MPCA structures of data, we represent the MPCA structure at a given interval by the cortege of projectors corresponding to MPCA at that interval; this representation has good and meaningful statistical properties which are discussed. To reveal underlying geometric structures that may be present in the data, clustering analysis of the PCA structures at various scales groups together scales with similar PCA structures and separate scales with dissimilar PCA structures.

For data with clear multiscale structures, the cluster analysis reveals some underlying structures in the data which conventional PCA cannot reveal due to the fact that such structures are obfuscated by other structures of higher variance. Each meaningful cluster corresponds to a structure in the data and we represent each cluster by the medoid point of the cluster and this representative is used to describe the structure of the data for cluster. We propose the Ratio of Distortion as a criterion for choosing an appropriate scale for MPCA for feature extraction especially for data with outliers and also as a criterion for choosing the PCA structure to describe each cluster in the clustering analysis of scales. Application of MPCA on artificial and real life examples shows that this can be useful. For data with multi-scale structures, the method was able to reveal some underlying structure in data. The method was particularly useful in mitigating the influence of outliers on the analysis of principal component without having to exclude such outliers explicitly.

# Appendix

**Table A1.** The angle between the original component and the result of the first principal component using MPCA at various scales for the data in example 3.

| | | Upper Limit | | | | | | | | |
|---|---|---|---|---|---|---|---|---|---|---|
| | SCALE | 1.0 | 0.9 | 0.8 | 0.7 | 0.6 | 0.5 | 0.4 | 0.3 | 0.2 | 0.1 |
| Lower Limit | 0.0 | 85.2543 | 6.6465 | 6.6465 | 6.6465 | 6.6465 | 6.6465 | 6.6465 | 6.4516 | 8.7675 | 14.9184 |
| | 0.1 | 85.2934 | 6.5879 | 6.5879 | 6.5879 | 6.5879 | 6.5879 | 6.5879 | 6.3704 | 8.5396 | 0.0000 |
| | 0.2 | 85.6238 | 6.1229 | 6.1229 | 6.1229 | 6.1229 | 6.1229 | 6.1229 | 5.6063 | 0.0000 | 0.0000 |
| | 0.3 | 86.0901 | 7.2010 | 7.2010 | 7.2010 | 7.2010 | 7.2010 | 7.2010 | 0.0000 | 0.0000 | 0.0000 |
| | 0.4 | 86.2657 | 90.0000 | 90.0000 | 90.0000 | 90.0000 | 90.0000 | 0.0000 | 0.0000 | 0.0000 | 0.0000 |
| | 0.5 | 86.2657 | 90.0000 | 90.0000 | 90.0000 | 90.0000 | 0.0000 | 0.0000 | 0.0000 | 0.0000 | 0.0000 |
| | 0.6 | 86.2657 | 90.0000 | 90.0000 | 90.0000 | 0.0000 | 0.0000 | 0.0000 | 0.0000 | 0.0000 | 0.0000 |
| | 0.7 | 86.2657 | 90.0000 | 90.0000 | 0.0000 | 0.0000 | 0.0000 | 0.0000 | 0.0000 | 0.0000 | 0.0000 |
| | 0.8 | 86.2657 | 90.0000 | 0.0000 | 0.0000 | 0.0000 | 0.0000 | 0.0000 | 0.0000 | 0.0000 | 0.0000 |
| | 0.9 | 86.2657 | 0.0000 | 0.0000 | 0.0000 | 0.0000 | 0.0000 | 0.0000 | 0.0000 | 0.0000 | 0.0000 |

*Note: The MPCA at scale 0-1 is the same as PCA. The cell for PCA as being marked with a grey-scale background*

**Table A2.** The ratio of distortion of MPCA at various scales for k = 2. This result is for the data given in example 3.

| | | Upper Limit | | | | | | | | |
|---|---|---|---|---|---|---|---|---|---|---|
| | SCALE | 1.0 | 0.9 | 0.8 | 0.7 | 0.6 | 0.5 | 0.4 | 0.3 | 0.2 | 0.1 |
| Lower Limit | 0.0 | 0.9030 | 1.0000 | 1.0000 | 1.0000 | 1.0000 | 1.0000 | 1.0000 | 1.0000 | 1.0000 | 1.0000 |
| | 0.1 | 0.9196 | 1.0000 | 1.0000 | 1.0000 | 1.0000 | 1.0000 | 1.0000 | 1.0000 | 1.0000 | 0.0000 |
| | 0.2 | 0.9734 | 1.0000 | 1.0000 | 1.0000 | 1.0000 | 1.0000 | 1.0000 | 1.0000 | 0.0000 | 0.0000 |
| | 0.3 | 0.9906 | 1.0000 | 1.0000 | 1.0000 | 1.0000 | 1.0000 | 1.0000 | 0.0000 | 0.0000 | 0.0000 |
| | 0.4 | 0.9971 | NaN | NaN | NaN | NaN | NaN | 0.0000 | 0.0000 | 0.0000 | 0.0000 |
| | 0.5 | 0.9971 | NaN | NaN | NaN | NaN | 0.0000 | 0.0000 | 0.0000 | 0.0000 | 0.0000 |
| | 0.6 | 0.9971 | NaN | NaN | NaN | 0.0000 | 0.0000 | 0.0000 | 0.0000 | 0.0000 | 0.0000 |
| | 0.7 | 0.9971 | NaN | NaN | 0.0000 | 0.0000 | 0.0000 | 0.0000 | 0.0000 | 0.0000 | 0.0000 |
| | 0.8 | 0.9971 | NaN | 0.0000 | 0.0000 | 0.0000 | 0.0000 | 0.0000 | 0.0000 | 0.0000 | 0.0000 |
| | 0.9 | 0.9971 | 0.0000 | 0.0000 | 0.0000 | 0.0000 | 0.0000 | 0.0000 | 0.0000 | 0.0000 | 0.0000 |

*Note: The MPCA at scale 0-1 is the same as PCA. The cell for PCA as being marked with a grey-scale background*

**Table A3.** The ratio of distortion of MPCA at various scales for k = 1. This result is for the data given in example 3.

| | | | | | Upper Limit | | | | | |
|---|---|---|---|---|---|---|---|---|---|---|
| SCALE | 1.0 | 0.9 | 0.8 | 0.7 | 0.6 | 0.5 | 0.4 | 0.3 | 0.2 | 0.1 |
| 0.0 | 0.4695 | 0.8297 | 0.8297 | 0.8297 | 0.8297 | 0.8297 | 0.8297 | 0.8139 | 0.7433 | 0.6748 |
| 0.1 | 0.4989 | 0.8511 | 0.8511 | 0.8511 | 0.8511 | 0.8511 | 0.8511 | 0.8360 | 0.7630 | 0.0000 |
| 0.2 | 0.6467 | 0.9341 | 0.9341 | 0.9341 | 0.9341 | 0.9341 | 0.9341 | 0.9279 | 0.0000 | 0.0000 |
| 0.3 | 0.8679 | 0.9532 | 0.9532 | 0.9532 | 0.9532 | 0.9532 | 0.9532 | 0.0000 | 0.0000 | 0.0000 |
| 0.4 | 0.9851 | NaN | NaN | NaN | NaN | NaN | 0.0000 | 0.0000 | 0.0000 | 0.0000 |
| 0.5 | 0.9851 | NaN | NaN | NaN | NaN | 0.0000 | 0.0000 | 0.0000 | 0.0000 | 0.0000 |
| 0.6 | 0.9851 | NaN | NaN | NaN | 0.0000 | 0.0000 | 0.0000 | 0.0000 | 0.0000 | 0.0000 |
| 0.7 | 0.9851 | NaN | NaN | 0.0000 | 0.0000 | 0.0000 | 0.0000 | 0.0000 | 0.0000 | 0.0000 |
| 0.8 | 0.9851 | NaN | 0.0000 | 0.0000 | 0.0000 | 0.0000 | 0.0000 | 0.0000 | 0.0000 | 0.0000 |
| 0.9 | 0.9851 | 0.0000 | 0.0000 | 0.0000 | 0.0000 | 0.0000 | 0.0000 | 0.0000 | 0.0000 | 0.0000 |

(Lower Limit labels the rows.)

*Note: The MPCA at scale 0-1 is the same as PCA. The cell for PCA as being marked with a grey-scale background*

**Table A4.** The percentage of pairwise distances excluded at various scales of MPCA. This result is for the data given in example 3.

| | | | | | Upper Limit | | | | | |
|---|---|---|---|---|---|---|---|---|---|---|
| SCALE | 1.0 | 0.9 | 0.8 | 0.7 | 0.6 | 0.5 | 0.4 | 0.3 | 0.2 | 0.1 |
| 0.0 | 0.00% | 11.20% | 11.20% | 11.20% | 11.20% | 11.20% | 11.20% | 15.85% | 34.71% | 73.53% |
| 0.1 | 26.47% | 37.68% | 37.68% | 37.68% | 37.68% | 37.68% | 37.68% | 42.32% | 61.18% | 0.00% |
| 0.2 | 65.29% | 76.50% | 76.50% | 76.50% | 76.50% | 76.50% | 76.50% | 81.15% | 0.00% | 0.00% |
| 0.3 | 84.15% | 95.35% | 95.35% | 95.35% | 95.35% | 95.35% | 95.35% | 0.00% | 0.00% | 0.00% |
| 0.4 | 88.80% | 100.00% | 100.00% | 100.00% | 100.00% | 100.00% | 0.00% | 0.00% | 0.00% | 0.00% |
| 0.5 | 88.80% | 100.00% | 100.00% | 100.00% | 100.00% | 0.00% | 0.00% | 0.00% | 0.00% | 0.00% |
| 0.6 | 88.80% | 100.00% | 100.00% | 100.00% | 0.00% | 0.00% | 0.00% | 0.00% | 0.00% | 0.00% |
| 0.7 | 88.80% | 100.00% | 100.00% | 0.00% | 0.00% | 0.00% | 0.00% | 0.00% | 0.00% | 0.00% |
| 0.8 | 88.80% | 100.00% | 0.00% | 0.00% | 0.00% | 0.00% | 0.00% | 0.00% | 0.00% | 0.00% |
| 0.9 | 88.80% | 0.00% | 0.00% | 0.00% | 0.00% | 0.00% | 0.00% | 0.00% | 0.00% | 0.00% |

(Lower Limit labels the rows.)